\documentclass[]{aa}
\usepackage{psfig,epsfig,graphicx}
\usepackage{amsmath}
\usepackage{natbib}
\bibpunct{(}{)}{;}{a}{}{,}

\begin{document}

\headnote{Research Note}
\title{Binary black holes in Mkns as sources of gravitational radiation for
space based  interferometers}
\author{F. De Paolis \inst{1}, G. Ingrosso\inst{1},
A.A. Nucita\inst{1},
 A.F. Zakharov\inst{2,3,4}
 }
\offprints{A.A. Nucita, \email{nucita@le.infn.it}} \institute{
Dipartimento di Fisica Universita di Lecce and INFN, Sezione di
Lecce,  Italy
 \and
Institute of Theoretical and Experimental Physics,
           25, B.Cheremushkinskaya st., Moscow, 117259, Russia
 \and
 Astro Space
Centre of Lebedev Physics Institute, Moscow,
\and  National
Astronomical Observatories, Chinese Academy of Sciences, 100012,
Beijing, China,
\\
}
\authorrunning{De Paolis et al.}
\titlerunning{Gravitational waves from massive binary black hole systems.}
\date{Received 30 April 2003 / accepted 8 August 2003}

\abstract{ The possibility that some Markarian objects (e.g. Mkn
501, Mkn 421 and Mkn 766) host  massive binary black hole systems
with eccentric orbits  at their centers has been considered. These
systems could be sources of gravitational radiation for
space-based gravitational wave interferometers like LISA and
ASTROD. In the framework of the Lincoln -- Will approximation we
simulate coalescence of such systems, calculate  gravitational
wave templates and discuss parameters of these binary black hole
systems corresponding to the facilities of LISA and ASTROD. We
discuss also the possibility to extract information about
parameters of the binary black hole systems (masses, of
components, distances between them, eccentricity and orbit
inclination angle with respect to line of sight)  from future
gravitational wave measurements.

}

\maketitle
\section{Introduction}

BL Lacertae objects, also known as Markarian objects (hereafter
Mkns), belong to the class of active galaxies according to the
well-established unified model on radio-loud active galactic
nuclei \citep{up}. These objects are thought to be dominated by
relativistic jets seen at small angles to the line of sight.

Until now, several astrophysical phenomena have been attributed to
binary black holes, like precession of jets \citep{bbr},
misalignment \citep{cw}, periodic outburst activity in the Quasar
OJ $287$ \citep{shv,lv} and precession of the accretion disk under
gravitational torque \citep{katz}.

It has been recently observed that some Mkns show a periodic
behavior in the radio, optical, X-ray and $\gamma$-ray light
curves that is possibly related to the presence of a massive
binary black hole with a jet along the line of sight or
interacting with an accretion disk \citep{yu}. Therefore, the
search for light curve variability, mainly in X-ray and
$\gamma$-ray wavelengths, can be considered as a method to probe
the existence of a massive binary black hole in the center of a
galaxy.

A question naturally arises: how can a binary system of massive
black holes be formed? The answer to this problem can be found in
the framework of the hierarchical vision of the universe
\citep{white}, for example if massive black hole systems form as a
result of merging processes between galaxies, each of them may
contain in the center a massive black hole \citep{rees,kr,rab}.
Recent observational signatures \footnote{The gravitational wave
spectrum from coalescing massive black hole binaries formed by
merging processes of their host galaxies has been studied by
\cite{Jaffe03}.} of such hypothesis were discussed, for example by
\cite{Yu01} who analyzed typical features of Fe $K_\alpha$ line
shapes.

At least three Mkns (i.e.  Mkn 501, Mkn 421 and Mkn 766) are
particularly well studied at high energies, revealing a possible
periodic behavior in their light curves.

Mkn 501, at $z=0.034$, shows a clear well-correlated 23 day
periodicity in $X$-ray and TeV energy bands with an observed TeV
flux ratio $f\simeq$ 8 between the maximum and minimum of the
signal \citep{pbf,hhi,kdk,nhc,kdk01}, while evidence for
correlations in the optical U-band is rather weak
\citep{cbb,aab_a,aab_b}. It has been also suggested that the
complex morphology of the jet and the peculiar behavior of its
spectral energy distribution are probably related to the presence
of a massive binary black hole \citep{cw,vr} at the center of Mkn
501.

Mkn 421, at $z=0.031$, is the brightest BL Lacertae object at
$X$-ray and UV wavelengths and it is the first extragalactic
source discovered at TeV energies \citep{pac}. This nearby source,
which has been recently observed by the XMM-Newton \citep{bsg} and
by Beppo-SAX \citep{mft} satellites, shows remarkable $X$-ray
variability correlated with strong activity at TeV energies
\citep{gwb} on a time-scale of $\simeq 10^{4}$ s \citep{mft} and
with a flux ratio $f \simeq 2$.

X-ray observations of the nearby Mkn 766, at $z=0.013$, have been
performed by the XMM-Newton satellite  \citep{bkt}. These
observations have revealed the presence of a strong $X$-ray
periodic signal with frequency $\simeq 2.4\times 10^{-4}$ Hz and
flux ratio $f\simeq 1.3$.

Based on the assumption that the periodic behavior of the observed
light curve for Mkn 501 is related to the presence of a binary
system of black holes (one of which emits a jet moving with
Lorentz factor $\gamma_{\rm b}$) on circular orbits, \cite{rm}
have proposed a method to determine the physical parameters of the
binary system from the observed quantities, i. e. signal
periodicity, flux ratio between maximum and minimum signal and
power law spectral index. However, binary black holes might be on
eccentric orbits and eccentricity values up to $0.8$--$0.9$ are
not necessarily too extreme \citep{fitchett}. This could happen
frequently if binary systems of black holes at galaxy centers form
generally as a consequence of merging processes between their host
galaxies.

Of course, due to gravitational wave emission, orbits tend, as a
first approximation, to circularize but this happens within a
time-scale of the same order of magnitude as the merging
time-scale \citep{Pet64,fitchett}. Therefore, if a massive binary
black hole is found at the center of a galaxy, it may happen that
the constituting black holes are still on eccentric orbits, in
which case the method proposed by \cite{rm} does not hold.

Hence, \cite{DePaolis02} studied the massive black hole binary
system possible in the center of some Mkn objects by assuming more
general elliptical orbits and considered the orbit eccentricity
$e$, the binary separation $a$ and Lorentz factor $\gamma_{\rm b}$
as free fit parameters used to determine the masses ($M_1$ and
$M_2$) of the two black holes from the observed $X$-ray
periodicity. Once the orbital parameters of the MBH (i.e. Massive
Black Hole) binaries are known, the values of the obtained orbital
separation, eccentricity and MBH masses are considered as initial
conditions in the time evolution of the binary systems.

In the present paper we study the evolution of the system and to
determine the gravitational wave waveforms, i.e. the amplitude of
the metric perturbation as a function of time. In doing this, we
simulate the evolution of binary black hole systems by using the
\citet{Lincoln90} approximation  and calculate gravitational wave
templates without any assumptions about the evolution of our
system on quasi-circular orbits. We also study the detectability
of the emitted gravitational waves by the next generation of
space-based interferometers like LISA \citep{Rein00} and ASTROD
\citep{Wu00,Ni02}.

The paper is structured as follows: in Section 2 we show how to
determine the massive black hole binary system parameters starting
from the observed $X$-ray periodicity toward the considered Mkn
objects. In Section 3, the model we use to simulate the binary
system evolution in the post$^{5/2}$-Newtonian approximation is
reviewed. In Section 4 we describe our results about the evolution
of binary system, profile gravitational wave templates, typical
times for the evolution of our binary systems before fitting into
the LISA frequency band. Finally, in Section 5 we draw some
conclusions.

\section{Initial conditions for a coalescing binary massive black hole system}

As discussed above, there exist a number of Mkn objects (like Mkn
501, Mkn 421 and Mkn 766) that show periodic activity in the
radio, optical, X-ray and $\gamma$-ray light curves. It has been
recently proposed that the observed periodicity is possibly
related to the presence of a massive binary black hole creating a
jet along the line of sight.

Following \citet{rm}, we assume that the observed signal
periodicity has a geometrical origin, being a consequence of a
Doppler-shifted modulation. It is therefore possible to relate the
observed signal period $P_{\rm obs}$ to the Keplerian orbital
period $P_{\rm k}$ by
\begin{equation}
P_{\rm obs}= (1+z) \left(1- \frac{v_{\rm z}}{c}\cos i
\right)P_{\rm k}~, \label{observedperiodicity}
\end{equation}
where $i\simeq 1/\gamma _{\rm  b}$ and $v_{\rm z}$ is the typical
jet velocity assumed to be $v_{\rm z}\simeq c(1-1/\gamma_{\rm
b}^2)^{1/2}$ \citep{spada}. Here, we rely  on the assumption that
the lighter black hole in the binary system is emitting a jet
which is moving toward the observer with Lorentz factor
$\gamma_{\rm b}$.
\begin{figure}[htbp]
\begin{center}
\vspace{11.7cm} \includegraphics{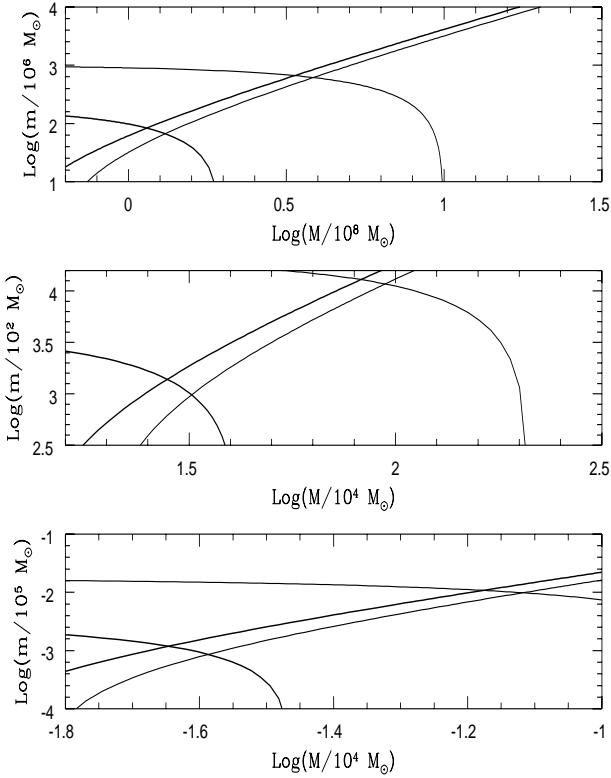} \caption{The secondary mass $m$
(in the text $M_1$) is shown as a function of the primary mass $M$
(in the text $M_2$) for the black hole binaries at the center of
Mkn 501, Mkn 421 and Mkn 766 (from the upper to the bottom panel).
The thick and thin lines represent the conditions expressed in
eqs. (\ref{massratio}) with Lorentz factor $\gamma _{\rm b}= 10$
and $\gamma _{\rm b}= 15$, respectively. The binary semi-major
axes $a$ are set to $5\times10^{16}$ cm, $2\times 10^{14}$ cm and
$1\times10^{13}$ cm for Mkn 501, Mkn 421 and Mkn 766. The orbital
eccentricity  in all cases is set equal to $0.5$. The intersection
between lines corresponding to the same Lorentz factor gives the
masses of the black holes in the considered binary system.}
\label{mcontromtotale}
\end{center}
\end{figure}

\begin{figure}[htbp]
\begin{center}
\vspace{7.cm} \includegraphics{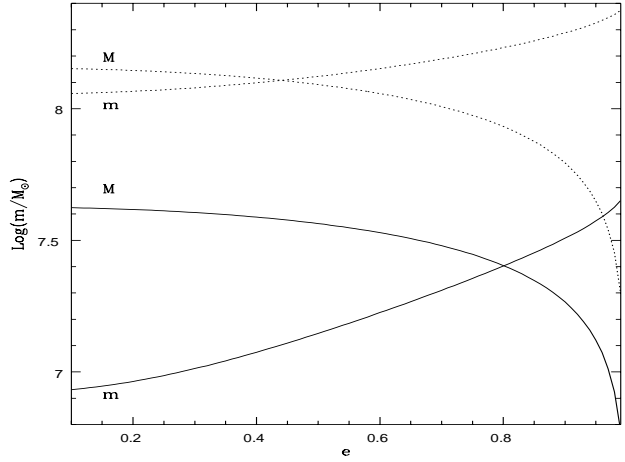} \caption{Primary and secondary black hole
masses are plotted as a function of the orbit eccentricity $e$ for
the binary system in Mkn 501.  Dotted lines correspond to Lorentz
factor $\gamma_{\rm b} = 10$ while solid lines correspond to
$\gamma_{\rm b}=30$. The semi-major axes is $a=8\times10^{16}$ cm.
Similar plots can be obtained for the Mkn 421 and Mkn 766.}
\label{mcontrogamma}
\end{center}
\end{figure}
The observed flux modulation due to Doppler boosting can be
written as
\begin{equation}
S(\nu) = \delta ^{3+\alpha}S^{\prime}(\nu)~, \label{fluxmkn}
\end{equation}
where $\alpha$ is the source spectral index \footnote{Values of
the power law index $\alpha$ are found to be 1.2, 1.7 and 2.11 for
Mkn 501, Mkn 421 and Mkn 766, respectively. For more details see
\citet{rm}, \citet{gvm} and \citet{bkt}.} and the Doppler factor
$\delta$ is given by
\begin{equation}
\delta = \frac{\sqrt{1-(v_{\rm z}^2+ v_{\rm ls}^2)/c^2}}{1-(v_{\rm
z} \cos i + v_{\rm ls} \sin i )/c}~. \label{doppler}
\end{equation}
Here, $v_{\rm z}$ is the jet velocity, $i$ is the inclination
angle between the jet axis and the line of sight and $v_{\rm ls}$
the component of the less massive black hole velocity along the
line of sight.

Depending on the position of $M_1$ along its orbit, the velocity
$v_{\rm ls}$ ranges between a minimum and a maximum value
corresponding, through eq. (\ref{doppler}), to the two extremal
values for the Doppler factor given by
\begin{eqnarray}
\delta_{\rm max}&=& \frac{\sqrt{1-\frac{(v_{\rm z}^2+v_{\rm
ls}^2)}{c^2}}} {1-\left(v_{\rm z} \cos i +\frac{2\pi R_{\rm a}
\sin i}{P_{\rm k} (1-e^2)^{1/2}}\right)c^{-1}}~,\nonumber \\ \\
\nonumber \delta_{\rm min}&=& \frac{\sqrt{1-\frac{(v_{\rm
z}^2+v_{\rm ls}^2)}{c^2}}} {1-\left(v_{\rm z} \cos i -\frac{2\pi
R_{\rm a} \sin i}{P_{\rm k} (1-e^2)^{1/2}}\right)c^{-1}}~,
\nonumber \label{deltamaxemin}
\end{eqnarray}
where $R_{\rm a}=M_2 a/(M_1+M_2)$, being $a$ and $e$ the binary
semi-major axis and the orbit eccentricity, respectively. From eq.
(\ref{fluxmkn}), the observed maximum and minimum fluxes modulated
by the Doppler effect turn out to be
\begin{eqnarray}
S_{\rm max}(\nu) &=& \delta_{\rm max} ^{3+\alpha} S^{\prime}(\nu)~,\nonumber \\
 \\
S_{\rm min}(\nu) &=& \delta_{\rm min} ^{3+\alpha}
S^{\prime}(\nu)~, \nonumber \label{flussi}
\end{eqnarray}
so that one obtains the condition $\delta _{\rm max}/ \delta_{\rm
min} \simeq f^{1/(3+\alpha)}$, where $f=S_{\rm max}(\nu)/S_{\rm
min}(\nu)$ is the observed maximum to minimum flux ratio.

By using eqs. (\ref{deltamaxemin}) and (\ref{flussi}) we have
\begin{equation}
R_{\rm a} = \frac{cP_{\rm k}}{2\pi}
\frac{f^{1/3+\alpha}-1}{f^{1/3+\alpha}+1}\left(\frac{1}{\sin i}
-\frac{v_{\rm z}}{c}\cot i \right)(1-e^2)^{\frac{1}{2}}~.
\label{omegar}
\end{equation}

Finally, eqs. (\ref{observedperiodicity}) and (\ref{omegar}) can
be rewritten in the following form (for more details see
\citet{DePaolis02})
\begin{equation} \left\{
\begin{array}{lll}
\displaystyle{\frac{M_2}{(M_1+M_2)^{2/3}}=
\frac{P_{\rm obs}^{1/3}}{[2\pi(1+z)G]^{1/3}}\frac{c}{\sin i}}\times \\ \\
~~~~~\displaystyle{
\frac{f^{1/3+\alpha}-1}{f^{1/3+\alpha}+1}\left(1
-\frac{v_{\rm z}}{c}\cos i \right)^{2/3}} (1-e^2)^{1/2},\\ \\
\displaystyle{M_1+M_2 =\left[\frac{2\pi(1+z)\left(1-\dfrac{v_{\rm
z}}{c}\cos i\right)}{P_{\rm obs}}\right]^2 \frac{a^3}{G}}~.
\end{array}
\right. \label{massratio}
\end{equation}
In these equations $\gamma_{\rm b}$, $e$ and $a$ have to be
considered as free model parameters to be determined by the
observed data ($P_{\rm obs}$, $f$ and $\alpha$) of the three Mkns
of interest. In this way, we are able to estimate the masses of
the black holes supposed to be composing each of the MBH binaries.
For the three considered Mkn objects, the secondary mass $M_1$ as
a function of the primary one $M_2$ is shown in Fig.
\ref{mcontromtotale}. The intersection between lines corresponding
to the same Lorentz factor ($\gamma_{\rm b} =10$ for the thick
lines and $\gamma_{\rm b} =15$ for the thin lines) gives the
masses of the black holes in the considered binary system.

In Fig. \ref{mcontrogamma} (for Mkn 501) the primary and secondary
black hole masses as a function of the orbit eccentricity
($\gamma_{\rm b}= 10, 30$ and $a = 8\times 10^{16}$ cm) are shown.
In this case, the black hole masses are in the range
$10^6-10^9~M_{\odot}$.

The previous method allows us to determine the orbital parameters
of the massive binary black hole possibly in the center of some
Mkn objects. The obtained values for the semi-major axes $a$ and
eccentricity $e$ obviously changes in time as a consequence of
emission of gravitational radiation. However, as shown by
\cite{Lincoln90}, when studying binary systems non-negligible
relativistic corrections in the equations of motion have to be
considered. It follows that in the simulation of the evolution of
the binary systems, the obtained values for the semi-major axes
and eccentricity have to be considered as the initial conditions
of the problem, i.e. $a_{\rm i}$ and $e_{\rm i}$, respectively.

\begin{figure*}[htbp]
$\begin{array}{c@{\hspace{1in}}c} \epsfxsize=3in
\epsffile{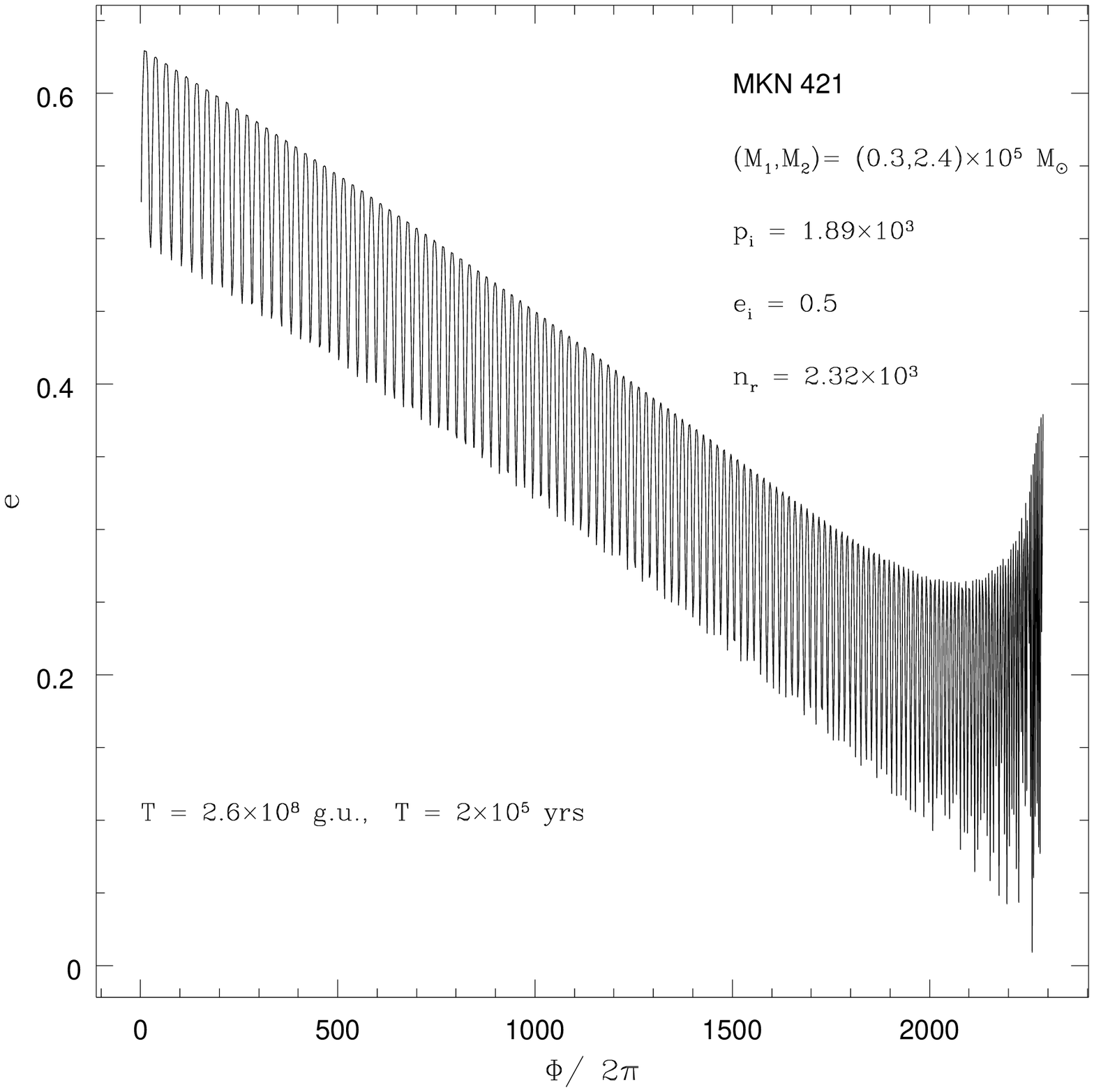} &
    \epsfxsize=3in
    \epsffile{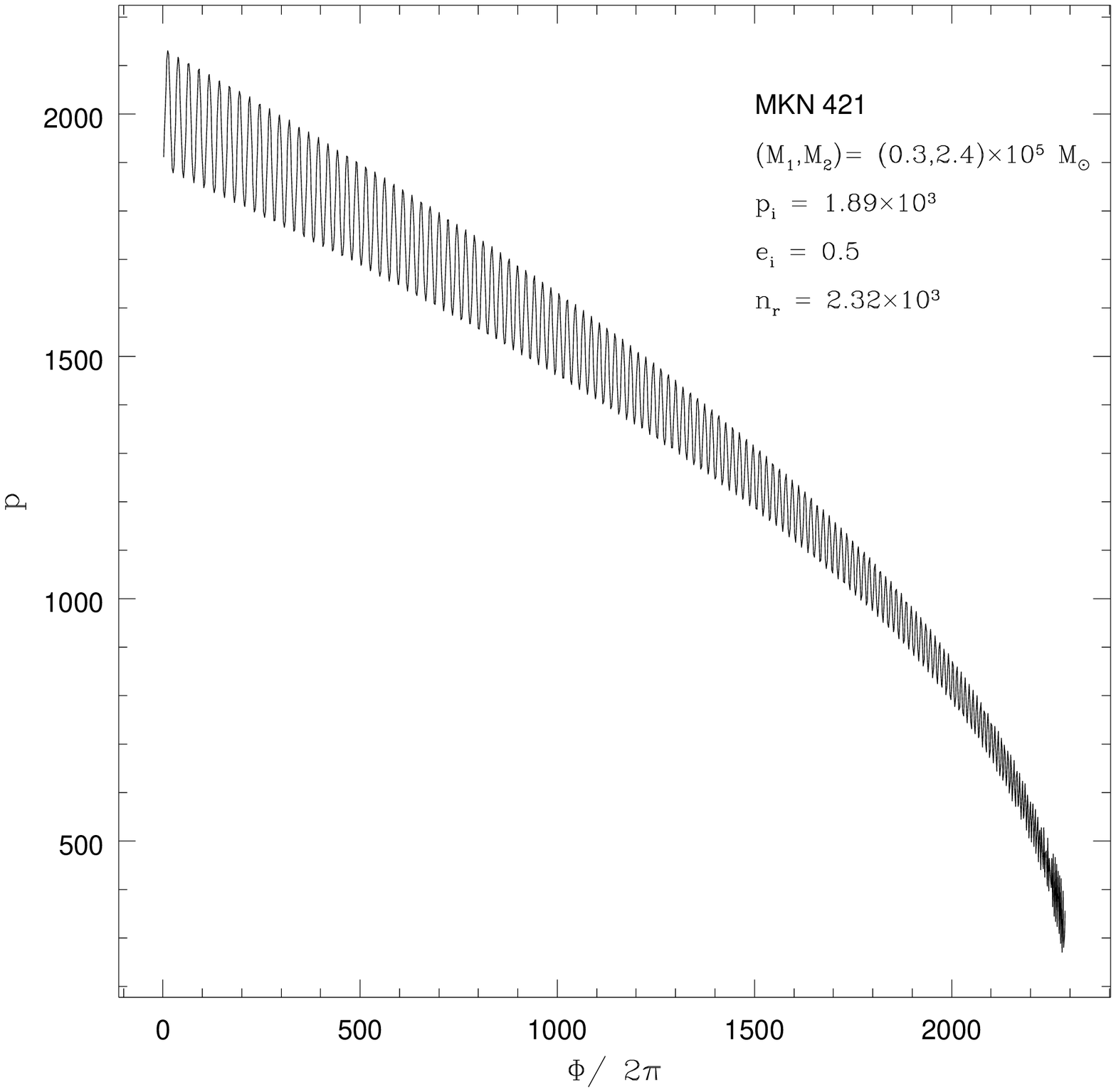} \\ [0.4cm]
\mbox{\bf a)} & \mbox{\bf b)}
\end{array}$
\caption{For Mkn 421,  the evolution of a MBH binary system is
shown. The orbital parameters reported in each panel allow to
reproduce the X-ray light curve periodicity. The evolutions of
both the MBH binary system eccentricity $a$) and  separation $b$)
are given as a function of the number of revolutions $\Phi/2\pi$.
The initial parameter $p_{\rm i}=a_{\rm i}(1-e_{\rm i}^2)$ is
given in unit of $GM/c^2$ being $M=M_1+M_2$.}
\label{evsrev+pvsrev}
\end{figure*}

\section{Model for a coalescing binary system evolution}

To analyze the binary black hole evolution we use a
(post)$^{5/2}$-Newtonian approximation developed by
\cite{Lincoln90}.\footnote{One can see also
\cite{Zakh96,Zakh97,Zakh98} where this approximation was used to
analyze the evolution of a binary neutron star system in which the
lower mass object is losing matter since it fills its Roche lobe.}

Earlier, \cite{Grish86} derived similar equations of motion for
isolated bodies in a (post)$^{5/2}$-Newtonian approximation taking
into account the effects of gravitational radiation friction which
lead to orbital shrink. In studying these effects \cite{Grish86}
derived their equations using osculating elements like those
reported in \cite{Lincoln90}. Here we use the \cite{Lincoln90}
approach since it is very convenient for our computational aims.
Of course, there are a lot of other more precise approaches which
may allow one to describe the final stages of evolution of binary
black hole systems (see, for example,
\cite{Buon99,Buon00,Fiziev01,Damour01,Buon02,Mora02,Gour02a,
Gour02b,OShau03,Buon03} and references therein). Even the
transition from inspiral stage to plunge for circular orbits was
discussed by \cite{Ori00} and for eccentric orbits by
\cite{OShau03}. \cite{Glamp02a,Glamp02b} used a Kerr metric
approximation to analyze the evolution of a binary black hole with
a small mass ratio, namely $m/M \sim 10^{-4} - 10^{-6},$ where $m$
and $M$ are the masses of the captured body and the central black
hole, respectively. Rapidly spinning massive black hole binary
systems as possible sources for LISA were considered by
\cite{Vecchio03} in the framework of the post$^{1.5}$--Newtonian
approximation and it was found that the black hole spin could be a
very essential factor and could drastically change the signature
of the gravitational wave signal. However, in this paper we
decided to use the Lincoln -- Will approximation since, as it was
mentioned before, $5/2$ is the first order of the post-Newtonian
approximation where one could introduce the gravitational
radiation friction in a self-consistent way.

Moreover, usually these binary black hole systems are so far from
the plunge stage of their evolution that it can be simulated by
using the lowest self-consistent post-Newtonian approximation.
Thus, one could use more simple treatments like that proposed by
\cite{Pet63,Pet64}. However, in the framework of these
approximations, the motion and the time evolution of the binary
system cannot be studied in detail since some relativistic
effects, such as the ``perihelion shift'' (see, for example,
\cite{Ehlers76,Grish86} for details), are not accounted for. In
addition, \cite{Ehlers76} have some doubts about the accuracy of
such methods since approaches similar to the \cite{Pet63} and
\cite{Pet64} approximations are mathematically inconsistent and
may lead to errors of the same order as the effects being
considered. Thus, for a more accurate treatment of the problem we
follow the \citet{Lincoln90} approach, to which we refer for more
details.

The orbital motion of two massive black holes moving around the
common center of mass is strongly influenced by gravitational
radiation losses. Viewing this as a Newtonian orbital motion with
perturbation suggests use of the osculating orbital element method
taken from celestial mechanics.
\begin{figure}[htbp]
\begin{center}
\vspace{9.0cm} \includegraphics{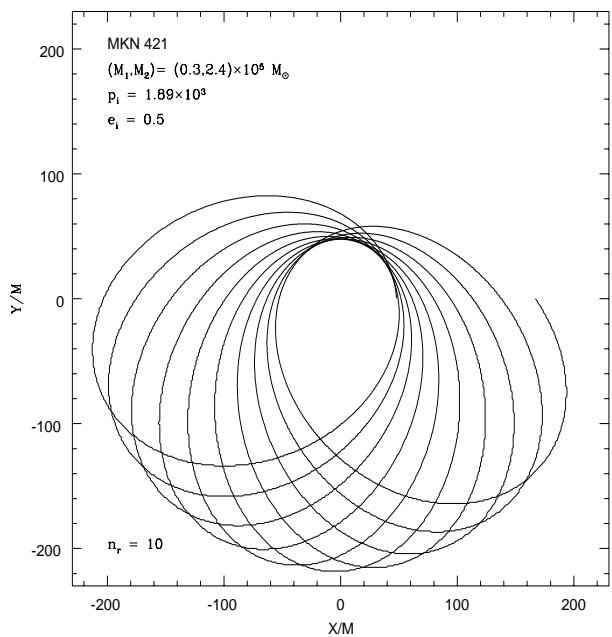} \caption{The orbital evolution of the black
hole binary at the center of Mkn 421 with initial parameters able
to reproduce the observed lightcurve periodicity is shown. The
orbital perihelion shift is evident.} \label{xvsy}
\end{center}
\end{figure}

The basic scenario is the following: at any time one can find a
Keplerian orbit which is tangent to the true orbit. This means
that both the position and the velocity of the particle on the
true orbit coincide with the position and the velocity of the
tangent Keplerian orbit at the considered time. Of course, at a
subsequent instant the actual orbit will be tangent to a different
Keplerian orbit. In the osculating orbit formalism a general
two-body orbit is generally described by six parameters: $i$ the
inclination of the orbit with respect to a reference plane,
$\Omega$ the line of the ascending node, $\omega$ the angle
between the line of node and the pericentric line, $a$ the
semi-major axis, $e$ the orbital eccentricity and $T$ the time of
pericentric passage. According to \cite{Lincoln90}, these
quantities are coupled by a set of first-order differential
equations, see eq. ($2.11a$)-($2.11c$) in the previously mentioned
paper, which can be numerically solved, by appropriately choosing
the initial conditions, obtaining $a=a(\Phi)$, $e=e(\Phi)$ and
$\omega=\omega(\Phi)$, $\Phi$ being the usual polar angle which in
turn depends on time $t$ through the relation $r^2
\dot{\Phi}=(Mp)^{1/2}$. Here, the auxiliary parameter $P$ is given
by $p=a(1-e^2)$.

Applying this approach, we are able to describe the motion and the
orbital evolution of the binary system. In Figs.
\ref{evsrev+pvsrev} $a$) and $b$) the evolution of the possible
Mkn 421 MBH binary system is shown. Notice that the orbital
parameters reported in the same panels correspond to a set of
values of the Lorentz factor $\gamma_{\rm b}$, semi-major axis $a$
and orbital eccentricity $e$ reproducing the observed X-ray light
curve periodicity. The evolution of both the MBH binary system
eccentricity $a$) and separation $b$) are given as a function of
the number of revolutions $\Phi/2\pi$. In the same panels the
expected coalescing time, calculated as in \cite{Lincoln90}, is
also reported both in geometrical units ($2.6\times 10^8$) and in
years ($2\times 10^5$ yrs) for the considered MBH binary system.
For comparison, the coalescing time scale of an eccentric binary
calculated using the Peters approximation (see eq. (5.14) in
\citet{Pet64}) gives $\simeq 1.3\times 10^5$ yrs. It is not
surprising that the two coalescing time scales differ, since in
\cite{Lincoln90} different approximations and relativistic effects
(as the perihelion shift) are taken into account. As evident, the
difference between the time scales evaluated by using these
methods is not dramatic, so that the \cite{Pet64} formalism can be
used, as a first approximation, to evaluate the coalescing time
for binaries that are (like ours) far enough from the plunge
stage.
\begin{figure*}[htbp]
\begin{center}
\vspace{11.0cm} \includegraphics{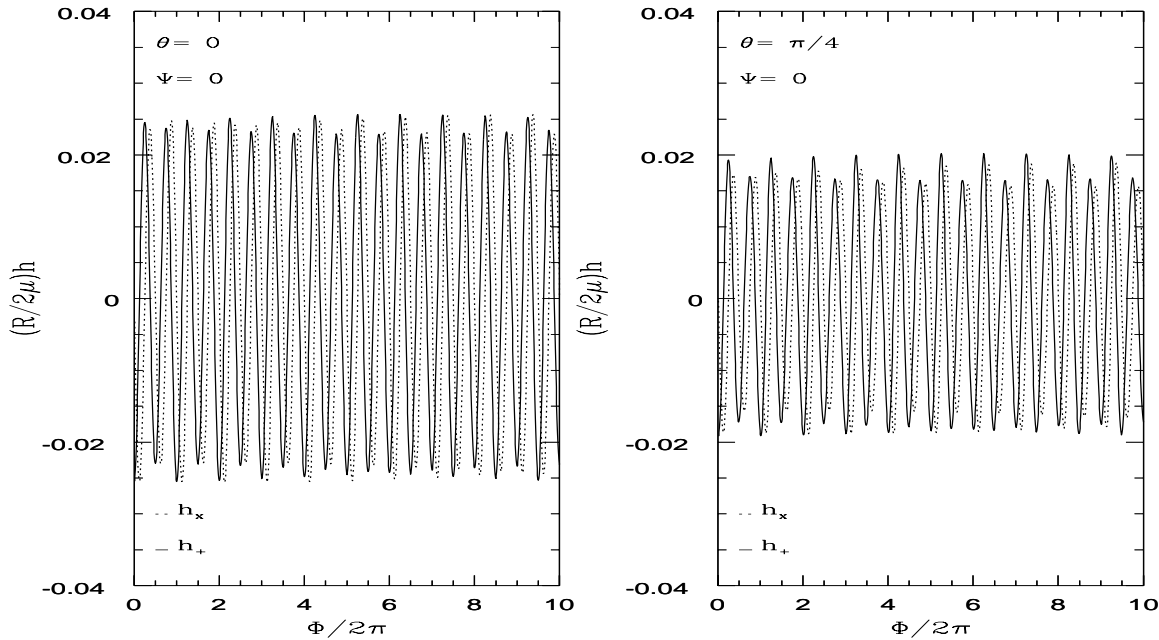} \caption{The two polarization states
($h_{\times}$ and $h_+$) of the gravitational waves emitted by the
MBH binary system (with reduced mass $\mu$), at distance $R$ from
Earth, for the selected model in Fig. \ref{evsrev+pvsrev} are
shown. The waveforms are given for the first 10 orbital
revolutions. The two parameters $\Theta$ and $\Psi$ determine the
direction of observation. In principle, the study of the
gravitational wave templates will allow one to extract information
about the MBH orbital parameters.} \label{fig500}
\end{center}
\end{figure*}
A further effect that can be accurately described by using the
\cite{Lincoln90} treatment is the orbital evolution of the black
hole binary system. For example in Fig. \ref{xvsy} we show the
orbital evolution of the black hole binary in Mkn 421 with initial
parameters chosen in order to reproduce the observed lightcurve
periodicity. The orbital perihelion shift is clearly evident.

Once the orbital motion of the binary system is known as a
function of time, the osculating orbital parameter formalism
allows us to determine the gravitational wave form, i.e. the
evolution in time of the metric perturbation. In particular, using
eqs. ($4.1a$)-($4.1b$) in \cite{Lincoln90}, we can evaluate the
polarization states $h_{\times}$ and $h_+$ of the gravitational
wave emitted by the considered MBH binary system. In Fig.
\ref{fig500} we plot, for the first $10$ orbital revolutions, the
expected wave forms depending on the $\Theta$ and $\Psi$
parameters which determine the observation direction.
\begin{figure}[htbp]
\begin{center}
\vspace{11.7cm} \includegraphics{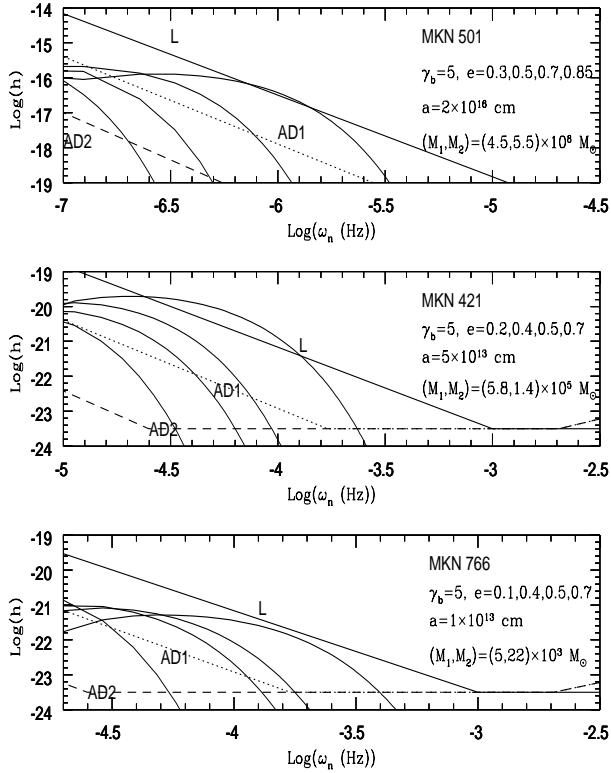}

\caption{Assuming particular values of the free model parameters
($\gamma_{\rm b}$, $e$ and $a$), we compare the expected
gravitational wave spectra with the LISA (L), ASTROD1 (AD1) and
ASTROD2 (ADF2) thresholds, respectively. Note that the values of
the black holes reported for each Mkn objects strongly depend on
the chosen parameters.} \label{gwspectra}
\end{center}
\end{figure}

\section{Discussion and conclusions}

Usually, at the first stage of evolution, our binary systems are
outside of the LISA frequency band since the typical frequency of
the emitted gravitational radiation is much lower than $10^{-4}$
Hz. However, other experiments, such as the ASTROD gravitational
wave detector \citep{Wu00,Ni02}, will have a much higher
sensitivity than LISA. In fact, even using the same laser power as
for LISA, the ASTROD sensitivity would be shifted to a frequency
lower by  a factor up to 60 (30 on average), \cite{Ni02}.
Moreover, as noted by \cite{Rud02}, if the LISA accelerometer
noise goal will be obtained, the ASTROD sensitivity at low
frequencies will be about 30 times better than that of LISA, as
indicated in Fig. 8 of \cite{Ni02}. \footnote{See also Fig. 13
from the paper by \cite{Rud02}.} Fig. \ref{gwspectra} shows the
possibility of detecting gravitational waves from such systems
using LISA and ASTROD(1) and ASTROD(2) (the detailed description
of these ASTROD facilities was given by \cite{Ni02}). Moreover, if
the absolute metrological accelerometer/intertial sensor can be
developed, there is even the possibility of reaching the ASTROD(3)
sensitivity curve with a shifting factor that could reach $10^{3}
- 10^{4}$ lower than LISA (see again Fig. 8 from the paper by
\cite{Ni02}). Note that ways to decrease the instrumental noise up
to $(1-3)\times 10^{-5}$ are discussed by \cite{Hughes02} (see
also \cite{Larson02}).

In principle there is even a non-negligible chance to determine
the inclination angle for a binary black hole system using
gravitational wave observations if we will have a possibility to
distinguish these templates for different $\Theta$ and $\Psi$
angles. Of course, it could be only a hypothetical chance to
extract this information from observations because one should
collect data for some years to reach the necessary sensitivity to
detect the emitted gravitational waves. The dependence of
gravitational wave templates on $\Theta$ and $\Psi$ angles could
be important to construct optimal filters for gravitational wave
detection.

\begin{acknowledgements}

AFZ would like to thank the Department of Physics of University of
Lecce, {\it INFN} - Sezione di Lecce, the National Astronomical
Observatories of Chinese Academy of Sciences for the hospitality
and profs. W.-T.~Ni, J.Wang and Dr. Z. Ma for very useful
discussions about the ASTROD project.

\end{acknowledgements}


\begin{thebibliography}{}

\bibitem[\protect\citeauthoryear{Aharonian, Akhperjanian, Barrio et al.}{1999a}]{aab_a}
Aharonian F., Akhperjanian A.G., Barrio J.A. et al., 1999a, A\&A
342, 69

\bibitem[\protect\citeauthoryear{Aharonian, Akhperjanian, Barrio et al.}{1999b}]{aab_b}
Aharonian F., Akhperjanian A.G., Barrio J.A. et al., 1999b, A\&A
349, 29

\bibitem[\protect\citeauthoryear{Begelman, Blandford \& Rees}{1980}]{bbr}
Begelman M.C.,  Blandford R.D. \&  Rees M.J. , 1980, Nat 287, 307

\bibitem[\protect\citeauthoryear{Boller, Keil, Tr\"{u}mper et al.}{2001}]{bkt}
Boller T., Keil R., Tr\"{u}mper J. et al., 2001, A\&A 365, 134

\bibitem[\protect\citeauthoryear{Brinkmann, Sembay, Griffiths et al.}{2001}]{bsg}
Brinkmann W., Sembay S., Griffiths R.G. et al., 2001, A\&A 365,
162

\bibitem[\protect\citeauthoryear{Buonanno}{2002}]{Buon02}
   Buonnanno, A. 2002, CQG 19, 1267

\bibitem[\protect\citeauthoryear{Buonanno, Chen \& Valisneri}{2003}]{Buon03}
   Buonnanno, A., Chen, Y. \& Valisneri,  M. 2003, PRD 65, 044020

\bibitem[\protect\citeauthoryear{Buonanno \& Damour}{1999}]{Buon99}
   Buonnanno, A. \& Damour,  T. 1999, PRD 59, 084006

\bibitem[\protect\citeauthoryear{Buonanno \& Damour}{2000}]{Buon00}
   Buonnanno, A. \& Damour,  T. 2000, PRD 62, 064015

\bibitem[\protect\citeauthoryear{Catanese, Bradbury, Beslin et al.}{1997}]{cbb}
Catanese M, Bradbury S.M., Beslin A.C. et al., 1997, ApJ 487, L143

\bibitem[\protect\citeauthoryear{Conway \& Wroble}{1995}]{cw}
Conway J.E. \&  Wroble J.M., 1995, ApJ 439, 98


\bibitem[\protect\citeauthoryear{Damour}{2001}]{Damour01}
    Damour,  T. 2001, PRD 64, 124013

\bibitem[\protect\citeauthoryear{De Paolis, Ingrosso \& Nucita}{2002}]{DePaolis02}
   De Paolis, F., Ingrosso G. \& Nucita A. 2002,  A\&A 388, 470

\bibitem[\protect\citeauthoryear{Ehlers et al.}{1976}]{Ehlers76}
   Ehlers, J.,  Rosenbum A., Goldberg J.N.  \& Havas , P. 1976, ApJ 208, L77

\bibitem[\protect\citeauthoryear{Fitchett}{1987}]{fitchett}
Fitchett M., 1987, MNRAS 224, 567

\bibitem[\protect\citeauthoryear{Fiziev \& Todorov}{2001}]{Fiziev01}
    Fiziev, P.P. \& Todorov, I. T. 2001, PRD 63, 104007

\bibitem[\protect\citeauthoryear{George, Warwick \& Bromage}{1988}]{gwb}
George I.M., Warwick R.S. \& Bromage G.E., 1988, MNRAS 232, 793

\bibitem[\protect\citeauthoryear{Glampedakis \& Kennefick}{2002}]{Glamp02a}
   Glampedakis, K. \& Kennefick, D. 2002, PRD 66, 044002

\bibitem[\protect\citeauthoryear{Glampedakis, Hughes \& Kennefick}{2002}]{Glamp02b}
   Glampedakis, K.  Hughes, S. \& Kennefick, D. 2002, PRD 66, 064005

\bibitem[\protect\citeauthoryear{Gourhoulhon, Grandclement \& Bonazzola}{2002a}]{Gour02a}
   Gourhoulhon, E., Grandclement, P. \& Bonazzola,  S.  2002a, PRD 65, 044020

\bibitem[\protect\citeauthoryear{Gourhoulhon, Grandclement \& Bonazzola}{2002b}]{Gour02b}
   Gourhoulhon, E., Grandclement, P. \& Bonazzola, S. 2002b,  PRD 65, 044021

\bibitem[\protect\citeauthoryear{Grishchuk \& Kopejkin}{1986}]{Grish86}
Grishchuk, L.P. \& Kopejkin, S.M.  1986, Equations of motion for
isolated bodies, in Relativity, Celestial Mechanics and
Astrometry, Proceedings of Symposium 114 of IAU. See also ``High
precision dynamical theories and observational verifications'',
Proceedings of the Symposium, Leningrad, USSR, May 28-31, 1985
(A87-24502 09-90), Dordrecht, D. Reidel Publishing Co., 1986, 19

\bibitem[\protect\citeauthoryear{Guainazzi, Vacanti, Malizia et al.}{1999}]{gvm}
Guainazzi M., Vacanti G. , Malizia A. et al., 1999, A\&A 342, 124

\bibitem[\protect\citeauthoryear{Hayashida, Hirasawa \& Ishikawa}{1998}]{hhi}
Hayashida N., Hirasawa H. , Ishikawa F. et
al., 1998, ApJ 504, L71

\bibitem[\protect\citeauthoryear{Jaffe \& Backer}{2003}]{Jaffe03}
Jaffe A.H. \& Backer D.C., 2003, ApJ 583, 616

\bibitem[\protect\citeauthoryear{Hughes}{2002}]{Hughes02}
   Hughes, S.A. 2002, MNRAS 331, 805

\bibitem[\protect\citeauthoryear{Katz}{1997}]{katz}
Katz J.I., 1997, ApJ 478, 527


\bibitem[\protect\citeauthoryear{Kormendy \& Richstone}{1995}]{kr}
Kormendy  J. \& Richstone D., 1995, ARA\&A 33, 581

\bibitem[\protect\citeauthoryear{Kranich, DeJager, Kestel et al.}{1999}]{kdk}
 Kranich, D., O.C. DeJager, O.C.,  Kestel, M. et al., 1999, in: {\it
Proc. of 26th International Cosmic Ray Conference (Salt Lake
City)} 3, p.358

\bibitem[\protect\citeauthoryear{Kranich, DeJager, Kestel et al.}{2001}]{kdk01}
 Kranich, D., O.C. DeJager, O.C.,  Kestel, M. et al., 2001, in: {\it
Proc. of 27th International Cosmic Ray Conference (Salt Lake
City)} 7, p. 2630

\bibitem[\protect\citeauthoryear{Larson, Hiscock \& Hellings}{2002}]{Larson02}
   Larson, S.L.,  Hiscock, W.A., Hellings R.W. 2002, PRD 62, 062001

\bibitem[\protect\citeauthoryear{Lehto \& Valtonen}
{1996}]{lv} Lehto  H.L. \&  Valtonen M.J., 1996, ApJ 460, 207

\bibitem[\protect\citeauthoryear{Lincoln \& Will}{1990}]{Lincoln90}
   Lincoln, C.W. \& Will C., 1990, PRD 42, 1123

\bibitem[\protect\citeauthoryear{Maraschi, Fossati, Tavecchio et al.}{1999}]{mft}
Maraschi L., Fossati G., Tavecchio F. et al., 1999,
Astropart.Phys. 11, 189

\bibitem[\protect\citeauthoryear{Mora \& Will}{2002}]{Mora02}
   Mora, T. \& Will C. 2002,  PRD 66, 101501(R)

\bibitem[\protect\citeauthoryear{Ni}{2002}]{Ni02}
Ni, W.-T. 2002, Inter. Journ. Mod. Phys. D 11, 947

\bibitem[\protect\citeauthoryear{Nishikawa, Hayashi, Chamoto et al.}{1999}]{nhc}
Nishikawa D., Hayashi S., Chamoto S. et al., 1999, in: {\it Proc.
of 26th International Cosmic Ray Conference (Salt Lake City)} 3,
354

\bibitem[\protect\citeauthoryear{Peters}{1964}]{Pet64}
   Peters, P.C.  1964, PR 136, 1124

\bibitem[\protect\citeauthoryear{Peters \& Mathews}{1963}]{Pet63}
   Peters, P.C. \& Mathews,  J. 1963, PR 131, 435

\bibitem[\protect\citeauthoryear{Protheroe, Bhat \& Fleury}{1998}]{pbf}
Protheroe R.J., Bhat C.L. ,Fleury  P. et al., 1998, in:{\it Proc.
25th International Cosmic Ray Conference (Durban)} 8, p 317

\bibitem[\protect\citeauthoryear{Punch, Akerlof, Cawley et
al.}{1992}]{pac} Punch M., Akerlof  C.W., Cawley M.F. et al.,
1998, Nat 358, 477

\bibitem[\protect\citeauthoryear{Ori \& Thorne}{2000}]{Ori00}
   Ori, A \& Thorne, K.S. 2000, PRD 62, 124022

\bibitem[\protect\citeauthoryear{O'Shaughnessy}{2003}]{OShau03}
   O'Shaughnessy, R. 2003, PRD 67, 044004

\bibitem[\protect\citeauthoryear{Rees}{1984}]{rees}
Rees M., 1984, ARA\&A  22, 471

\bibitem[\protect\citeauthoryear{Reinhard}{2000}]{Rein00}
Reinhard, R. 2000,   LISA -- Detecting and Observing Gravitational
Waves, in ESA Bulletin 103, 33

\bibitem[\protect\citeauthoryear{Richstone, Ajhar \& Bender}{1998}]{rab}
Richstone D., Ajhar E.A.\& Bender R., 1998, Nat 395, 14


\bibitem[\protect\citeauthoryear{Rieger \& Mannheim }{2000}]{rm}
Rieger F.M. \& Mannheim K., 2000, A\&A 359, 948

\bibitem[\protect\citeauthoryear{Rieger \& Mannheim }{2003}]{rm3}
Rieger F.M. \& Mannheim K., 2003, A\&A 397, 121

\bibitem[\protect\citeauthoryear{R\"udiger}{2002}]{Rud02}
R\"udiger, A. 2002,  Inter. Journ. Mod. Phys. D 11, 963

\bibitem[\protect\citeauthoryear{Sillanp\"{a}\"{a}, Haarala, Valtonen et al.} {1988}]{shv}
Sillanp\"{a}\"{a} A., Haarala S., Valtonen M.J. et al., 1988, ApJ
325, 628

\bibitem[\protect\citeauthoryear{Spada}{1999}]{spada}
Spada M., 1999, Astropart. Phys. 11, 59

\bibitem[\protect\citeauthoryear{Urry \& Padovani}{1995}]{up}
Urry L.M. \& Padovani P., 1995, PASP 107, 803

\bibitem[\protect\citeauthoryear{Vecchio}{2003}]{Vecchio03}
Vecchio A., 2003, preprint astro-ph/0304051

\bibitem[\protect\citeauthoryear{Villata \& Raiteri}{1999}]{vr}
Villata M. \& Raiteri C.M., 1999, ApJ 347, 30

\bibitem[\protect\citeauthoryear{White}{1997}]{white}
White S.D.M., in: B\"{o}rner G., Gottl\"{o}ber S. (eds), {\it The
evolution of the Universe: Report of the Dahlem Workshop} (Berlin,
1997), p. 227

\bibitem[\protect\citeauthoryear{Wu, Xu \& Ni }{2000}]{Wu00}
Wu A.-N., Xu X.H. \&   Ni, W.-T. 2000, Inter. Journ. Mod. Phys. D
9, 201

\bibitem[\protect\citeauthoryear{Yu}{2002}]{yu}
Yu Q., 2002, MNRAS, 331, 805; preprint astro-ph/0109530

\bibitem[\protect\citeauthoryear{Yu \& Lu}{2001}]{Yu01}
Yu Q. \& Lu Y., 2001, A\&A 377, 17

\bibitem[\protect\citeauthoryear{Zakharov}{1996}]{Zakh96}
Zakharov, A.F. 1996, Astron. Rep. 40,  552

\bibitem[\protect\citeauthoryear{Zakharov}{1997}]{Zakh97}
Zakharov, A.F. 1997, A\&SS 252,  213

\bibitem[\protect\citeauthoryear{Zakharov}{1998}]{Zakh98}
Zakharov, A.F. 1998, Supernovae as Source of Gravitational
Radiation, in Gravitational Waves, Second Edoardo Amaldi
Conference held in CERN, Switzerland, 1-4 July 1997. Ed. E.
Coccia, G. Veneziano, and G. Pizzella. World Scientific, 247
\end{thebibliography}
\end{document}